\documentclass[aps,prb,
twocolumn,superscriptaddress,amsmath,amssymb,verbatim]{revtex4-2}
\usepackage{amsfonts,stmaryrd,wasysym,graphicx,multirow,textcomp,subfigure,makecell}
\usepackage[font=small,labelfont=bf, justification=justified, format=plain]{caption} 

\usepackage{url}

\usepackage{dcolumn}
\usepackage{bm}
\usepackage[colorlinks=true,citecolor=blue,urlcolor=blue]{hyperref}
\hypersetup{colorlinks=true, urlcolor=blue, citecolor=cyan, pdfborder={0 0 0},}
\usepackage{xcolor}

\def\be{\begin{equation}}
        \def\ee{\end{equation}}
\def\bea{\begin{eqnarray}}
        \def\eea{\end{eqnarray}}

\renewcommand{\Im}{\mathrm{Im}\,}

\newcommand{\p}{\partial}

\renewcommand{\Im}{\mathrm{Im}\,}

\newcommand{\bs}{\boldsymbol}
\DeclareMathAlphabet{\bi}{OML}{cmm}{b}{it}

\def\ba{\begin{aligned}}
\def\ea{\end{aligned}}
\def\be{\begin{equation}}
\def\ee{\end{equation}}
\def\bearr{\begin{eqnarray}}
\def\eearr{\end{eqnarray}}
\def\la{\langle}
\def\ra{\rangle}
\def\l{\left}
\def\r{\right}

\begin{document}

\title{Non-linear in-plane spin current in spin-orbit coupled 2D hole gases}

\author{Srijan Chatterjee}
\thanks{Corresponding author: \href{mailto:srijanc24@iitk.ac.in}{srijanc24@iitk.ac.in}}
\affiliation{Department of Physics, Indian Institute of Technology Kanpur, Kanpur 208 016, India}

\author{Tarun Kanti Ghosh}
\affiliation{Department of Physics, Indian Institute of Technology Kanpur, Kanpur 208 016, India}

\date{\today}

\begin{abstract}
The non-linear transport of charge and spin due to the emergence of band geometric effects has garnered much interest in recent years. In this work, we show that a linear in-plane spin current vanishes, whereas a non-linear (second-order) in-plane spin current exists for a generic two-dimensional system having time-reversal symmetry. The intrinsic second-order spin current originates from the spin Berry curvature polarizability. The formulation when applied to 2D hole gases with the $k^3$ Rashba spin-orbit coupling reveals the existence of both transverse and longitudinal second-order spin currents normal to the spin orientation. Interestingly, anisotropic spin-orbit couplings can generate collinearly polarized spin current (spins polarized in the direction of spin current) in the second-order. The effects of anisotropy are explored by introducing an additional Dresselhaus spin-orbit coupling and electromagnetic radiation over the isotropic Rashba system. The generation and control over the multiple in-plane spin currents may have important applications in spintronic devices.
\end{abstract}

\maketitle

\section{Introduction}
The development of efficient spintronic devices involve the generation and effective manipulation of spin currents \cite{Intro_1,spintronics_book,Fabian,SHE-review}.
A spin current refers to the transport of spin angular momentum, with or without an accompanying charge transport. 
In addition to the spatial degrees of freedom of the charge carriers, the spin current also enjoys a spin degree of freedom, which leads to the following classification. 
In two-dimensional (2D) systems, the spin current is spatially restricted to the 2D plane, while the electronic spin operator has three components $(\hat S_x,\hat S_y,\hat S_z)$, each having a pair of eigenstates. 
The spins can either be polarized perpendicular or parallel to the plane of transport, which leads to `out-of-plane' and `in-plane' spin currents respectively. 
Among the in-plane spin currents, the spins can be oriented along the flow (collinearly polarized spin currents) or perpendicular to it.
A spin current that involves in principle, zero Joule heating and a zero net charge flow is called a `pure' spin current. The primary source of pure spin current in 2D spin-orbit coupled systems is the phenomenon of spin Hall effect (SHE).    

Spin Hall Effect belongs to the family of anomalous Hall effects, where a transport transverse to an applied electric field is observed in the absence of an external magnetic field\cite{SHE-review}. It involves a transverse transport of spins, such that carriers in different spin states (up and down) 
are accumulated on the opposite edges of the sample. This was first studied by D'yakonov and Perel \cite{dyakonov1971current} and later by Hirsch \cite{hirsch1999spin}. 
The key difference in the mechanisms of SHE and anomalous charge Hall transport lies in the breaking 
of time-reversal symmetry ($\mathcal{T}$). The latter is mainly observed in magnetic materials where 
$\mathcal{T}$ is broken by the spontaneous magnetization, whereas SHE does not require $\mathcal{T}$ 
to break. The central role in SHE is played by spin-orbit coupling (SOC), which arises due 
to a broken structural inversion symmetry ($\mathcal{P}$) of the material.
There are namely the Rashba spin-orbit coupling \cite{bychkov1984oscillatory} and the Dresselhaus spin-orbit 
coupling \cite{dresselhaus1955spin}. 
This article, like most others \cite{schliemann2003nonballistic},\cite{schliemann2003anisotropic} will 
involve the interplay of these interactions in the system. In the context of SHE, spin currents can either be transverse or longitudinal to the applied electric field. The longitudinal spin current is attributed to the anisotropic Rashba-Dresselhaus SOC which gives rise to a transverse charge current \cite{transverse_charge_current}, which in turn leads to a spin current in the longitudinal direction due to SHE \cite{conserved_spin_current, long-SHE}.

Based on the mechanism, SHE can be classified as intrinsic \cite{sinova2004universal} and extrinsic \cite{extrinsic}. The latter occurs due to spin-dependent scattering processes from impurities like skew-scattering \cite{SMIT1955877,SMIT195839} and side-jumping \cite{sidejump}. The intrinsic SHE, 
on the other hand has its origin in the geometry of the quantum states of the charge carriers, manifested 
in the Berry curvature. Although the theory of anomalous Hall effect was given by Karplus and Luttinger back in 1954 \cite{karplus1954hall}, it was only after the discovery of Berry phase in 1984 \cite{berry1984quantal} when the role of quantum geometry in Hall transport phenomena began to be explored \cite{Murakami_BC}.

Understanding the role of quantum geometry in the generation of intrinsic spin current is of prime importance in current times 
when non-linear anomalous Hall effect has already been attributed to the quantum metric \cite{AmitAgarwal, quantum_metric, Kaplan, nonlinear-qnuantum_metric} 
and the Berry curvature dipole \cite{Sodemann_Fu,PhysRevB.104.115140}. 
Berry connection polarizability (BCP), which is the corrected Berry connection due 
to an applied electric field is known to give rise to non-linear intrinsic Hall responses \cite{Third_order_Hall}
in $\mathcal{PT}$ symmetric anti-ferromagnets \cite{BCP1,BCP2}.

Attempts to explain the intrinsic spin current in terms of quantum geometrical elements are also underway. 
In Ref. \cite{PKapri}, it is shown that the intrinsic spin Hall current can be derived from the 
anomalous velocity of electrons in $\mathcal{T}$-broken Rashba spin-orbit coupled systems, thus emphasizing the role of the Berry curvature.
Further, instead of the conventional Kubo formula used in \cite{sinova2004universal} and elsewhere to calculate the spin current, Zhang et al. \cite{zhang2024intrinsic} shows that a spin Berry curvature (SBC) can be 
derived from a simple perturbation theory to calculate the spin Hall conductivity. 
Coming to non-linear SHE, apart from the Drude effect \cite{Drude,PhysRevB.104.115140} and the spin Berry curvature dipole \cite{SBCD}, which lead to extrinsic SHE, Zhang et al. \cite{zhang2024intrinsic} using their perturbative approach, have derived an intrinsic second-order spin current from the spin Berry curvature polarizability (SBCP). 
A similar semi-classical derivation has been done in Ref. \cite{Hui_Wang}, while Ref. \cite{noncentrosymmetric} uses the Boltzmann transport theory to obtain a second-order response to the applied electric field.
Ref. \cite{sayansarkar} provides a comprehensive survey of various intrinsic and extrinsic effects leading to non-linear SHE.

In this paper, we apply the perturbative formalism developed in \cite{zhang2024intrinsic} to 
two-dimensional hole gases (2DHG) formed at the hetero-junctions of III-V p-doped 
semiconductors to calculate the linear and second-order spin conductivities. 
Previously, studies on linear SHE in hole gases were carried out by \cite{murakami2003dissipationless,Bernavig,schliemann2005spin,HongLiu,conserved_spin_current,SHE_hole_expt}. 
The calculations are done in the $\mathcal{T}$ symmetric heavy hole band of the spectrum, 
discussed in detail by Murakami et al. \cite{murakami2003dissipationless} and Schliemann and Loss \cite{schliemann2005spin}. It is known that the linear spin conductivity in hole gases is contributed mainly by the intrinsic mechanism, since the vertex corrections due to impurity scattering vanish for such systems \cite{Bernavig,Murakami}.
We particularly emphasize the spin orientations in the SBC and SBCP tensors and 
note that the former is out-of-plane while the latter can have in-plane components. 
This is shown to be true for any two-level system where $\mathcal{T}$ is preserved \cite{classification}. 
Further, the role of anisotropy in the dispersion relations of these systems is also discussed. Apart from the Dresselhaus effect, the role of electromagnetic radiation in introducing anisotropy is also explored. 

This paper is arranged in the following way: section(\ref{sec:II}) introduces the formalism of 
second-order spin current and applies it to a general two-level $\mathcal{T}$ symmetric system. 
Section(\ref{secIII}) presents the results obtained for a 2DHG with the Rashba and Dresselhaus SOC. Section(\ref{secIV}) discusses the effect of radiation on the first and second-order spin conductivity of 
a Rashba spin-orbit coupled hole gas. 
Finally, we conclude with discussions and potential applications of our results in the field of spintronics.


\section{\label{sec:II} The formalism}
\subsection{Second-order spin current using perturbation theory}\label{sec:IIA}
In this section, we discuss the general formalism of linear and nonlinear spin currents
resulting from SBC and SBCP, in $\mathcal{T}$ preserved systems. This formalism 
is based on the Boltzmann transport equation within relaxation time approximation and the second-order correction to the Bloch states due to external electric field \cite{zhang2024intrinsic}.

The conventional spin current operator is defined 
as $ \hat{j}^{l}_{i} = \hbar(\hat S_{l} \hat v_{i} + \hat v_{i} \hat S_l)/2$,
where $ \hat v_{i} $ is the band velocity operator and 
$\hat S_l$ is a generalized spin operator.
 
The expectation value of the spin current operator in the single-particle states $\vert n \ra $ 
of the $n$-th band is given by
$ j^{l}_{i,n} = \langle n|\hat{j}^{l}_{i}|n\rangle $.

The total spin current from $n$ bands in two-dimensional systems can be evaluated as
\begin{equation}\label{total spin current}
    J^{l}_{i} = \sum_n\int_{k} [d{\bf k}] ~ j^{l}_{i,n} f_n({\bf k}),
\end{equation}
where $ [d{\bf k}] = (d^2k)/(2\pi)^2 $, the subscript $i$ ($i=x,y$) and the superscript 
$l$ ($l=x,y,z$) denote the 
directions of the spin current and spin polarization of the charge carrier, respectively. 
Also, $f_n({\bf k})$ is the non-equilibrium Fermi-Dirac occupation number for 
the $n$-th energy band. 

An in-plane homogeneous electric field ${\bf E} = E_x \hat {\bf x} + E_y \hat {\bf y} $ is applied
to the system, due to which  

the perturbation to the Hamiltonian is
$H^\prime = e {\bf E} \cdot {\bf r}$, where $e$ is the charge of a carrier. 
The normalized perturbed state up to second-order in electric field is $\vert \tilde n \rangle = \vert n^{(0)} \rangle + \vert n^{(1)} \rangle + \vert n ^{(2)} \rangle$ \cite{Sakurai, bransdenquantum}.

Here $\vert n^{(0)} \rangle$ is an unperturbed eigenstate, 
\be \label{1st-order}
\vert n^{(1)} \ra = \sum_{m \neq n } \frac{- e {\bf E} \cdot {\bf A}_{mn} }
{\varepsilon_m^{(0)} - \varepsilon_n^{(0)} }
\vert m^{(0)}\ra,
\ee
and 

\begin{widetext}
\begin{align} \label{2nd-order-state}
 |n^{(2)}\rangle = e^2\left[\sum_{m\neq n} \left[\sum_{p\neq n}\frac{({\bf E\cdot A}_{mp})({\bf E\cdot A}_{pn})}{\left(\varepsilon^{(0)}_n-\varepsilon^{(0)}_p \right) \left (\varepsilon^{(0)}_n-\varepsilon^{(0)}_m \right)} 
 - \frac{({\bf E\cdot A}_{mn})({\bf E\cdot A}_{nn})}{\left(\varepsilon^{(0)}_n-\varepsilon^{(0)}_m\right)^2}\right] |m^{(0)}\rangle - \frac{1}{2}\sum_{p \neq n} \frac{\vert({\bf E\cdot A}_{np})\vert^2}{\left(\varepsilon^{(0)}_n-\varepsilon^{(0)}_p\right)^2} \vert n^{(0)}\rangle\right]
\end{align}
\end{widetext}
are the first- and second-order corrections, respectively. 
${\bf A}_{mn} =  \la m^{(0)} \vert i {\bs \nabla}_{\bf k} \vert n^{(0)} \ra $ is the inter-band Berry connection and $\varepsilon_n^{(0)}$ are the unperturbed energy eigenvalues.

The electric field induced modified spin current is
$ \tilde j^{l}_{i,n} = \langle \tilde n |\hat{j}^{l}_{i}| \tilde n\rangle $.
After simplification, one can obtain (up to quadratic order in $ {\bf E}$)

\begin{equation}\label{modified-spin-current}
    \tilde{j}^{l}_{i,n} = j^{l,(0)}_{i,n} - e\Omega^{l}_{ij,n}E_{j} + e^{2}\Pi^{l}_{ijk,n}E_{j}E_{k}, 
\end{equation}
where $j^{l,(0)}_{i,n} = \langle n^{(0)}|\hat{j}^{l}_{i}\vert n^{(0)}\rangle$, $ \Omega^{l}_{ij,n}$ is called the spin Berry curvature and $\Pi^{l}_{ijk,n}$ is called the spin Berry curvature polarizability.
 The derivation of SBC and SBCP for a two-level system is given in Appendix (\ref{App:2nd-order}).

Within the relaxation time approximation \cite{Ashcroft76}, the non-equilibrium Fermi-Dirac occupation
number $f_n({\bf k}) $ can be obtained by solving the Boltzmann transport equation
\be
 \frac{e}{\hbar}{\bf E} \cdot {\bs \nabla}_{\bf k} f_n({\bf k}) = 
 - \frac{[f_n({\bf k}) - f_{n}^{(0)}({\bf k}) ]}{\tau},
\ee
where $ \tau $ is the relaxation time of the charge carriers and 
$f_n^{(0)}({\bf k})= 1/(e^{[\epsilon_n({\bf k})-\mu]/(k_BT)} +1)$ is the occupation number in absence 
of the electric field.
The iterative solution for $f_n({\bf k}) $ can be written as
\begin{equation}
f_n({\bf k}) = \sum_{p=0,1,2,...}\left(\frac{e\tau}{\hbar}\right)^{p} 
\frac{\partial^{p}f_n^{(0)}}{\partial k^{p}_{i}}E^{p}_{i}.
\end{equation}

The total modified spin current is given by
\begin{equation}
    \tilde J^{l}_{i} = \sum_{n}\sum_{p}\int_{k} [d{\bf k}] \tilde{j}^{l}_{i,n}\left(\frac{e\tau}{\hbar}\right)^{p}\frac{\partial^{p}f_n^{(0)}}{\partial k^{p}_{i}}E^{p}_{i}.
\end{equation}
Now putting the expression for $\tilde{j}^{l}_{i,n}$ from Eq. (\ref{modified-spin-current}), we get
\begin{equation}\label{non-linear spin current}
   \tilde J^{l}_{i} = \sigma^{l}_{i} + \chi^{l}_{ij}E_{j} + \Gamma^{l}_{ijk}E_{j}E_{k} + ... \:,
\end{equation} 
where the spin transport coefficients are given by
\begin{align}
    \sigma^{l}_{i} &= \sum_n \int_{k} [d{\bf k} ] j^{l,(0)}_{i,n} f_n^{(0)},\\
    \label{linear-spin conductivity}\chi^{l}_{ij} &= e \sum_n \int_{k} [d{\bf k} ] \Big( j^{l,(0)}_{i,n} \frac{\tau}{\hbar}
    \frac{\partial f_n^{(0)}}{\partial k_{j}} - \Omega^{l}_{ij,n}f_n^{(0)}\Big),
\end{align} 
\begin{align}
\label{2nd-order-spin conductivity}
\Gamma^{l}_{ij\kappa} = e^{2}  \sum_n \int_{k} [d{\bf k}] \Big( j^{l,(0)}_{i,n}\frac{\tau^{2}}
{\hbar^{2}}\frac{\partial^{2} }{\partial k_{j}\partial{k}_{\kappa}} &- 
\frac{\tau}{\hbar}\Omega^{l}_{ij,n}\frac{\partial  }{\partial k_{\kappa}} \notag \\
 &+ \Pi^{l}_{ij\kappa,n} \Big) f_n^{(0)} 
\end{align}

These are the expressions of the zeroth-order, first-order, and second-order 
spin conductivities, respectively. 
The $\tau$-dependent and $\tau$-independent terms in the non-linear spin conductivities 
are the extrinsic and intrinsic contributions, respectively.
For $\mathcal{T}$-symmetric systems, the term involving odd power of $\tau$ will vanish exactly,
while the even power of $\tau$ will have a finite contribution \cite{zhang2024intrinsic}.
Therefore, the $\tau$-linear term  in Eqs. (\ref{linear-spin conductivity}) and
(\ref{2nd-order-spin conductivity}) will be absent for $\mathcal{T}$-symmetric systems, where the structural inversion ($\mathcal{P}$) symmetry is broken. The spin currents computed in the following sections will have components both transverse and longitudinal to the applied electric field. The name spin `Hall' current/conductivity is reserved only for the transverse components.


\subsection{Results for a generic 2-level $\mathcal{T}$-symmetric system}\label{sec:IIB}
Let us consider a general time-reversal symmetric two-level system whose Hamiltonian is given by
\begin{equation}
    H = \frac{\hbar^2k^2}{2m}\mathbb{I} + {\bf d}({\bf k})\cdot {\bs \sigma}, 
\end{equation}
where $\mathbb{I}$ is the $2\times 2$ identity matrix and ${\bf d}=\big(d_x({\bf k}),\;d_y({\bf k})\big)$ and 
${\bs \sigma} = (\sigma_x,\sigma_y)$ are the Pauli matrices. Apart from the kinetic energy, we have 
\begin{equation}\label{2-level_Ham}
   H_s= {\bf d}\cdot{\bs \sigma }=d_x\sigma_x + d_y\sigma_y.
\end{equation}
Since there is no mass term containing $\sigma_z$, $\mathcal{T}$ is preserved. 
Before proceeding further, we note that spin is not a conserved quantity here, since the Hamiltonian in Eq. (\ref{2-level_Ham}) does not commute with the spin operators, defined below. This is generically true for any spin-orbit coupled system.
Further, the conventional spin current, defined in section(\ref{sec:IIA}), is also not conserved and is strongly influenced by spin precession and relaxation\cite{Intro_1,spintronics_book,Fabian}. 
This issue has been addressed by an alternative definition of spin current\cite{Niu-proper-spin-current}, which yields a `conserved' spin current under certain conditions. However, for weak spin-orbit coupling and non-linear spin transport, the conventional definition is still widely used\cite{Drude,sayansarkar,zhang2024intrinsic,Hui_Wang}. A detailed comparison between the two definitions of spin current can be found in Ref.\cite{properdefinitionspincurrent} the supplementary material of Ref.\cite{sayansarkar}.  

The eigenvalues and eigen states of $H_s$ are obtained as functions of ${\bf k}$:
\begin{equation}\label{eigen}
    \varepsilon_{\pm} ({\bf k}) = \pm\sqrt{d_x^2 + d_y^2}, ~~~
   |\pm\rangle = \frac{1}{\sqrt{2}}\begin{pmatrix}1 \\ \pm e^{i\psi} \end{pmatrix},
\end{equation}  
where $\psi=\tan^{-1}(d_y/d_x)$. 
Next, the Pauli matrices can be written in the eigen-basis of $H_s$ as
\begin{align}
  \label{sigma_xy} 
  \tilde \sigma_x & =  
  \begin{pmatrix} 
  \cos\psi & -i\sin\psi \\ 
  i\sin\psi &  -\cos\psi 
  \end{pmatrix}, 
  \tilde \sigma_y =  
  \begin{pmatrix} 
  \sin\psi & i\cos\psi \\ 
  -i\cos\psi &  -\sin\psi 
  \end{pmatrix},\\
   \label {sigma_z}
  \tilde \sigma_z  & =  
  \begin{pmatrix} 0 & 1 \\ 
  1 &  0 
  \end{pmatrix}. 
\end{align}
The band velocity components are given by

\begin{equation}\label{velocity}
    \hat{v}_i = \frac{\hbar k_i}{m}\mathbb{I} + \frac{1}{\hbar}\frac{\partial d_j}{\partial k_i}\sigma_j. 
\end{equation}

For the sake of generality, the spin operators are considered to be $2\times2$ Hermitian matrices, 
which may be functions of ${\bf k}$ \cite{Effective_spins}. They can be expanded in terms of the usual Pauli matrices and 
the identity matrix as 
\begin{equation}
   \hat S_l = \sum_{\lambda=0,x,y,z} B_\lambda({\bf k}) \sigma_\lambda, 
\end{equation}
with $\sigma_0 = \mathbb{I} $.
Using this definition of spin operators the spin current operator ($\hat{j}^{l}_i$) gets the following form
\begin{equation}
    \hat{j}^l_i = \frac{\hbar k_i}{m}\sum_{\lambda=0,x,y,z} B_\lambda({\bf k}) \sigma_\lambda 
    + \frac{1}{\hbar}\frac{\partial d_j}{\partial k_i}B_j({\bf k}).
\end{equation}
The spin current operator can have either in-plane ($l=x,y$) or out-of-plane ($l=z$) components.

The expressions for the SBC components (see Appendix (\ref{App:2nd-order})) for this two-level system are given by
\begin{equation}\label{sbc2}
    \Omega^l_{ij,\pm} = \frac{2\hbar}{\varepsilon_g^2}\ \Im{\Big[\langle \pm |\hat{j}^{l}_{i}|\mp\rangle\langle \mp|\hat{v}_{j}|\pm\rangle}\Big],
\end{equation}
In Eq. (\ref{sbc2}) we find that the expression of spin Berry curvature 
involves the off-diagonal matrix elements of the spin current and velocity operators in the $|\pm\rangle$ basis.
If we assume that the components of the in-plane spin current operators depend only on $\sigma_x$ and $\sigma_y$, (which is the case for the heavy hole gas in  (\ref{secIII})) their off-diagonal elements are purely imaginary according to Eq. (\ref{sigma_xy}) and their product will be real. Hence
\begin{equation}
    \Omega^{l}_{ij} = 0,\:\:\:\:\: l=x,y.
\end{equation}
This shows that the first-order spin current for such a system is strictly out-of-plane. 
Therefore, we wish to study the non-linear (2nd-order) in-plane spin current, due to the SBCP components (see Appendix (\ref{App:2nd-order})) given by
\begin{equation}\label{sbcp2}
    \Pi^{l}_{ijk,\pm} =  \frac{\hbar^2}{\varepsilon_g^4}  \langle \mp|\hat v_{j}|\pm\rangle \langle \pm|\hat v_{k}|\mp\rangle \left(\langle \mp|\hat{j}^{l}_{i}\vert\mp\rangle -\langle \pm|\hat{j}^{l}_{i}|\pm\rangle\right) .
\end{equation}

Further, it is should be noted that for the out-of-plane spin current, the SBCP components vanish. This is because the diagonal elements of $\sigma_z$ are zero in the $\vert \pm\rangle$ basis (Eq. (\ref{sigma_z})), making $\langle \pm\vert \hat j^z_i\vert\pm\rangle =0 $.
Hence,
\begin{equation}
    \Pi^z_{ijk} = 0.
\end{equation}

From Eq. (\ref{non-linear spin current}), ignoring the background spin current \cite{background_spin_current}, we can write
\begin{align}
    &\tilde{J}^z_i = \chi^z_{ij}E_j\:, \\
    &\tilde{J}^{l}_{i} = \Gamma^{l}_{ijk}E_jE_k\:, \:\:\:\:\: l=x,y.
\end{align}

In the subsequent sections, this formalism has been applied to spin-orbit coupled 2DHG, 
where the components of ${\bf d}({\bf k})$ are determined by the forms 
of the spin-orbit interactions. 


\section{ Heavy hole gas with Rashba and Dresselhaus spin-orbit coupling }\label{secIII}

A 2D hole gas is created by applying a confining potential transverse to a plane, thereby trapping the particles on it. Since the holes are formed in the valence band, their total angular momentum is $J =3/2,1/2$ ($J = L+S,\: L=1,S=1/2$). The angular momentum state $J=3/2$ is four-fold degenerate and the confining potential lifts this degeneracy partially, splitting states with $J_z=\pm3/2$ (heavy holes) and $J_z=\pm1/2$ (light holes) \cite{Winkler}. 
The heavy hole and light hole bands further split due to the inversion asymmetry induced Rashba or Dresselhaus spin-orbit coupling. 
The energy gap between the heavy and light hole bands depends inversely on the width of the confining potential well ($L_z$). If $L_z$ is made vanishingly small, the energy gap becomes extremely large and at low energies only the heavy hole band gets filled. 
In this approximation therefore, the heavy holes describe a two-level system independent of the other bands, with `effective spin' $\hat S_z=\pm \frac{3\hbar}{2}\sigma_z$. However, the in-plane effective spins $\hat S_x, \hat S_y$ vanish instead of reducing to Pauli matrices \cite{Bulaev_Loss, HH-LH_mixing}.\\  
For a finite $L_z$, the in-plane effective spin operators can be obtained by treating the mixing between the heavy hole and light hole bands perturbatively away from ${\bf k}=0$ \cite{Effective_spins}.

\begin{figure}[htbp]
\centering
\includegraphics[trim=0cm 0cm 0cm 0cm, clip, width=1\linewidth]{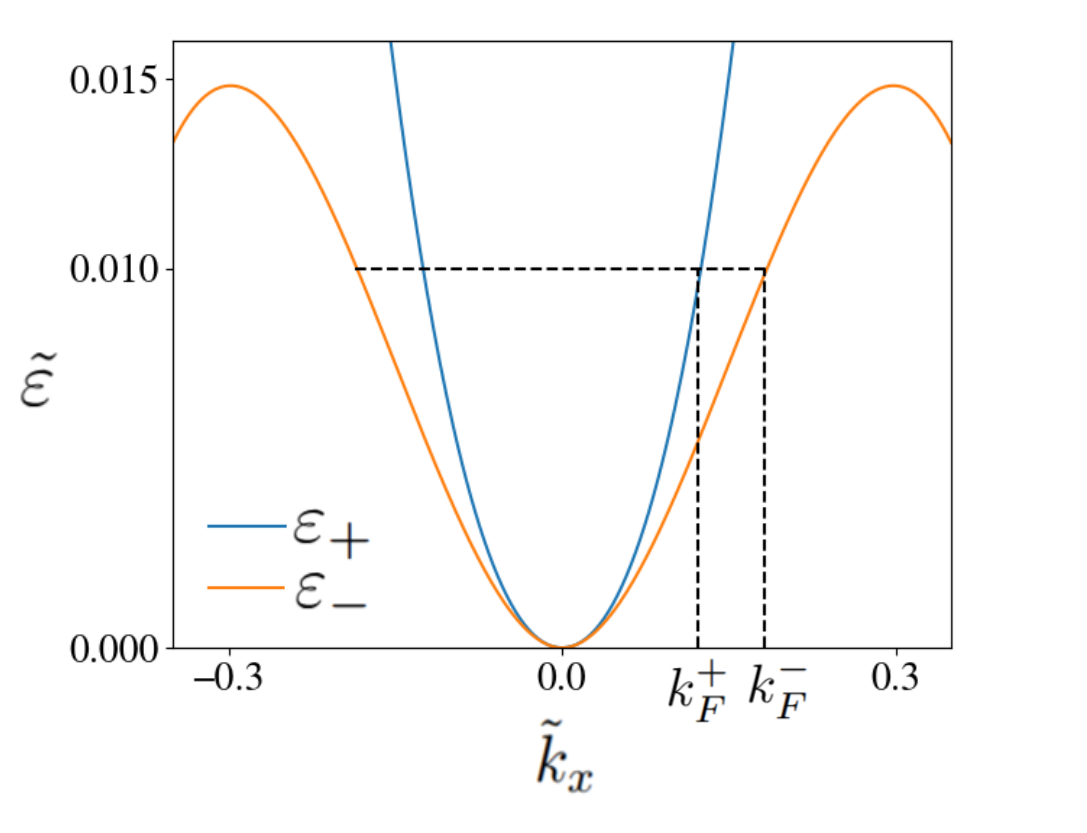} 
\captionsetup{format=plain, font=small, labelfont=bf, justification = raggedright}
\caption{Two spin-split energy bands versus $ \tilde k_x $ for $\beta/\alpha = 0.5$ are shown. Here $k_F^{\pm} $ are 
for a given Fermi energy.}
\label{fig:disp}
\end{figure}

An effective Hamiltonian of a heavy hole with $k$-cubic Rashba and Dresselhaus
spin-orbit couplings formed at the p-type III-V semiconductor heterojunctions 
can be written as \cite{Winkler,Ojasvi, Bulaev_Loss, Bashab_Dey} 
\begin{equation}
H = \frac{\hbar^{2}k^{2}}{2m} \mathbb{I} +  \l[ i\alpha k_{-}^{3} \sigma_+ - 
\beta k_{-}k_{+} k_{-}\sigma_{+} + h.c \r].
\end{equation}
Here $m$ is the effective mass of a heavy hole and, $\alpha$ and $\beta$ are the coupling strengths of 
the Rashba and Dresselhaus interactions respectively. Also, 
$k_{\pm}=k_{x}\pm i k_{y}$, $\sigma_{\pm} = (\sigma_x\pm i \sigma_y)/2 $ and $h. c.$ stands for 
the Hermitian conjugate.  The energy dispersion relation is given by
\begin{equation*}
    {\varepsilon}_n ({\bf k}) = \frac{\hbar^2 k^2}{2m} + 
    n k^{2}\sqrt{(\alpha k_{x}-\beta k_{y})^{2} + (\alpha k_{y}-\beta k_{x})^{2}},
\end{equation*}
where $n= \pm $ denote the two dispersive bands. They have been plotted in Fig. (\ref{fig:disp}).
The corresponding normalized eigen-spinors are given by
\begin{equation}\label{15}
    |n \rangle = |\pm\rangle = \frac{1}{\sqrt{2}}\begin{pmatrix}1 \\ \pm e^{i\psi} \end{pmatrix},
\end{equation}
where the $\alpha $ and $\beta $ dependent phase angle $\psi$ is given by 
$\psi= 2\phi-\phi^{\prime}$ with $\phi = \tan^{-1}(k_{y}/k_{x})$ and 
$\phi^{\prime}= \tan^{-1}\left[(\alpha k_{x}-\beta k_{y})/(\alpha k_{y}-\beta k_{x}\right)]$.
The energy band gap, $\varepsilon_g({\bf k}) = \varepsilon_+({\bf k}) - \varepsilon_-({\bf k}) $, is given by,
\begin{equation}
    \varepsilon_g({\bf k}) = 2k^{2}\sqrt{(\alpha k_{x}-\beta k_{y})^{2} + (\alpha k_{y}-\beta k_{x})^{2}}.
\end{equation}
It can be seen from the band gap expression that there is a line degeneracy along the
$k_y=k_x$ line for $\alpha=\beta $. On the other hand, for $\alpha=\beta$, the band gap is maximum 
($\varepsilon_g^{\text{max} }= 2 \alpha k^3$) along the $k_y=-k_x $ line.
Fig. (\ref{fig:disp}) also shows that out of the two bands, the lower one has a local maximum which turns out to be at $k = \hbar^2/3m\alpha$ in the pure Rashba model \cite{schliemann2005spin}. Since the energy decreases monotonically beyond this maximum, we get an upper cut-off momentum for this low-energy model.

The momentum ($ {\bf k}  $) dependent effective spin operators for the $k$-cubic Rashba system are given by \cite{Effective_spins}
\begin{align}
    &\hat S_x({\bf k} )  = \hbar\left[-S_0 k_y\mathbb{I} + S_1\left[(k_x^2-k_y^2)\sigma_x + 2k_xk_y\sigma_y\right]\right], \label{sx_modified}\\
    &\hat S_y ( {\bf k} ) = \hbar\left[S_0 k_x\mathbb{I} + S_1\left[(k_x^2-k_y^2)\sigma_y - 2k_xk_y\sigma_x\right]\right] \label{sy_modified},\\
    &\hat S_z({\bf k}) = \frac{3\hbar}{2}\sigma_z, \label{sz_modified}
\end{align}
where $S_0, S_1$ are system dependent parameters. They are given by 
$ S_0 = \alpha m_e/(\hbar^2 \gamma_2) $ and 
\begin{align}
    &S_1 = \left[\frac{3}{4\pi^2} - \frac{256\gamma_2^2}{3\pi^4(3\gamma_1 + 10\gamma_2)^2}\right]L_z^2,
\end{align}
where $m_e$  being the electron's rest mass and
$\gamma_1, \gamma_2$ are the Luttinger parameters \cite{Luttinger_first_paper}.
It can be easily seen that these effective spin operators do not satisfy the canonical commutation relations but they transform under time-reversal as $ \mathcal{T}^\dagger \hat S_l(- {\bf k} ) \mathcal{T} = - \hat S_l ({\bf k} ) $ with $\mathcal{T} $ being the time-reversal operator for spin-1/2 particles. 
Thus, the strong 2D confinement and the low energy limit gives rise to heavy holes with an out-of-plane pseudo-spin $\hat S_z$ and in-plane pseudo-spins $\hat S_x,\hat S_y$ which behave like spin-$1/2$ quasiparticles.

Since the effective spin operators are derived for a Rashba-dominated spin-orbit coupled system\cite{Effective_spins}, $\beta/\alpha$ is chosen small enough to safely use these spin operators when calculating the in-plane spin current. 
The Dresselhaus coupling is generally weak as compared to the Rashba coupling. Further, the strength of the Rashba coupling can be controlled using gate voltages \cite{Gate_control_soc}. For the out-of-plane spin current, the spin operator is simply $\sigma_z$ and $\beta/\alpha$ can be chosen greater than $1$ as well.

In the following subsections, the first- and second-order intrinsic spin conductivity are calculated by numerically integrating the SBC and SBCP components over the $k$-space. To facilitate the numerics, the dispersion relation and all the velocity and spin current components are calculated in polar coordinates $(k,\phi)$.
We have introduced a dimensionless wave-vector $\tilde k = k/k_h$ with $k_h = \hbar^2/(m\alpha)$ and
the dimensionless energy $\tilde \varepsilon = \varepsilon/\varepsilon_h $ with
$\varepsilon_h = \alpha k_h^3$.
The parameters $S_0$ and $S_1$ get scaled as $\tilde S_0 = S_0k_h$ and $\tilde S_1=S_1k_h^2$, their typical values chosen from \cite{Effective_spins}. The other parameters used  are
$\alpha = 0.2\: \text{eVnm}^3$ and effective mass $ m = 0.41 m_e$.


\subsection{Spin Berry curvature and out-of-plane linear spin conductivity}
The linear spin conductivity is calculated using the spin Berry curvature given in Eq.(\ref{sbc2}). 

The out-of-plane, transverse component of the SBC ($\Omega^z_{yx}$) due to an electric field along the $x$ direction is given by,
\be \label{sbc-rd}
\Omega^z_{yx,\pm} = \mp \frac{3\sin\phi}{4k_h^2\tilde{k}^3}
\l[ \frac{3  \cos\phi^\prime - 2r(2\sin\psi - \sin\phi^\prime)}
{1 + r^2 - 2r\sin(2\phi)} \r],
\ee
with the anisotropic parameter $r = \beta/\alpha $.

\begin{figure}
\centering
\includegraphics[trim=0cm 0cm 1cm 2cm, clip, width=0.8\linewidth]{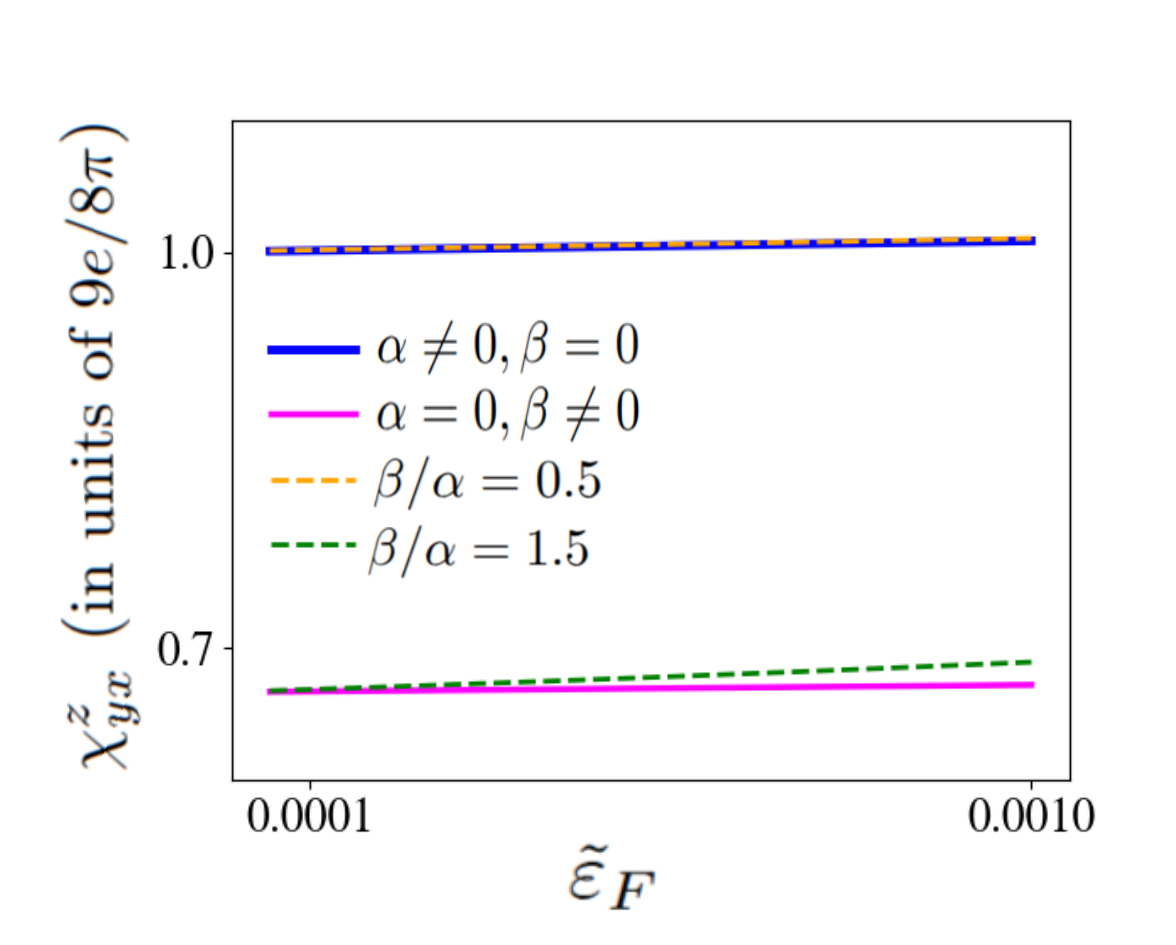}
\captionsetup{format=plain, font=small, labelfont=bf, justification = raggedright}
\caption{Linear spin Hall conductivity of the hole gas versus the
scaled Fermi energy ($\varepsilon/\varepsilon_h$) for different values of $\beta/\alpha$.}
\label{fig:linear_spin conductivity}
\end{figure}

Using Eqs.(\ref{linear-spin conductivity}) and (\ref{sbc-rd}), we reproduce the known result
for the intrinsic linear spin Hall conductivity
in the pure Rashba case ($r=0 $), 
which is  $\chi^{z,R}_{yx} \simeq 9e/(8\pi)$ \cite{schliemann2005spin}. 
On the other hand, in the pure Dresselhaus case ($ \alpha=0, \beta \neq 0)  $, 
we get 
$ \chi_{yx}^{z,D} \simeq 6e/(8\pi)$.
Therefore, $ \chi_{yx}^{z,D} $ is reduced by a factor of 2/3 as compared to 
$ \chi_{yx}^{z,R} $.
The linear spin Hall conductivity $\chi^z_{yx}$ is plotted as a function of the Fermi energy for 
different values of $r=\beta/\alpha$ in Fig. (\ref{fig:linear_spin conductivity}).\\ 
The longitudinal component of the SBC is given by,
\be \label{long}
\Omega^z_{xx,\pm} = \mp \frac{3\cos\phi}{4k_h^2\tilde{k}^3}
\l[ \frac{3  \cos\phi^\prime - 2r(2\sin\psi - \sin\phi^\prime)}
{1 + r^2 - 2r\sin(2\phi)} \r].
\ee
The longitudinal spin conductivity $\chi^z_{xx}$ is zero in the pure Rashba and pure Dresselhaus cases. However, it is non-zero in the presence of both Rashba and Dresselhaus interactions. This happens because the anisotropic Rashba-Dresselhaus SOC gives rise to a transverse charge current \cite{transverse_charge_current}, which in turn leads to a spin current in the longitudinal direction due to SHE \cite{conserved_spin_current}.
For $r = 0.5$, $\chi^z_{xx}\simeq 0.3331 (9e/8\pi)$ and for $r = 1.5$, $\chi^z_{xx}\simeq 0.2199 (9e/8\pi)$. The spin conductivities vary only marginally with the Fermi energy. 
These results are in contrast to the ones obtained for electron gases \cite{long-SHE, S.Q.Shen}.  

\subsection{Spin Berry curvature polarizability and in-plane second-order spin conductivity}
 Since the first-order spin current is polarized out-of-plane, 
 the first non-zero contribution from the in-plane spin current comes at the second-order. 
 The in-plane components of the SBCP tensor are calculated according to Eq. (\ref{sbcp2}). 
 The electric field in the $x$ direction fixes the last two indices of the SBCP tensor 
 at `$x$'. These indices may be dropped for a simpler notation. Thus, we have four components ($\Pi^{x}_{x},\Pi^{x}_{y},\Pi^{y}_{x},\Pi^{y}_{y}$) with the spin being oriented either along the $x$ or $y$ direction.

\begin{figure}
    \includegraphics[trim=1cm 0cm 0cm 0cm, clip, width=\linewidth]{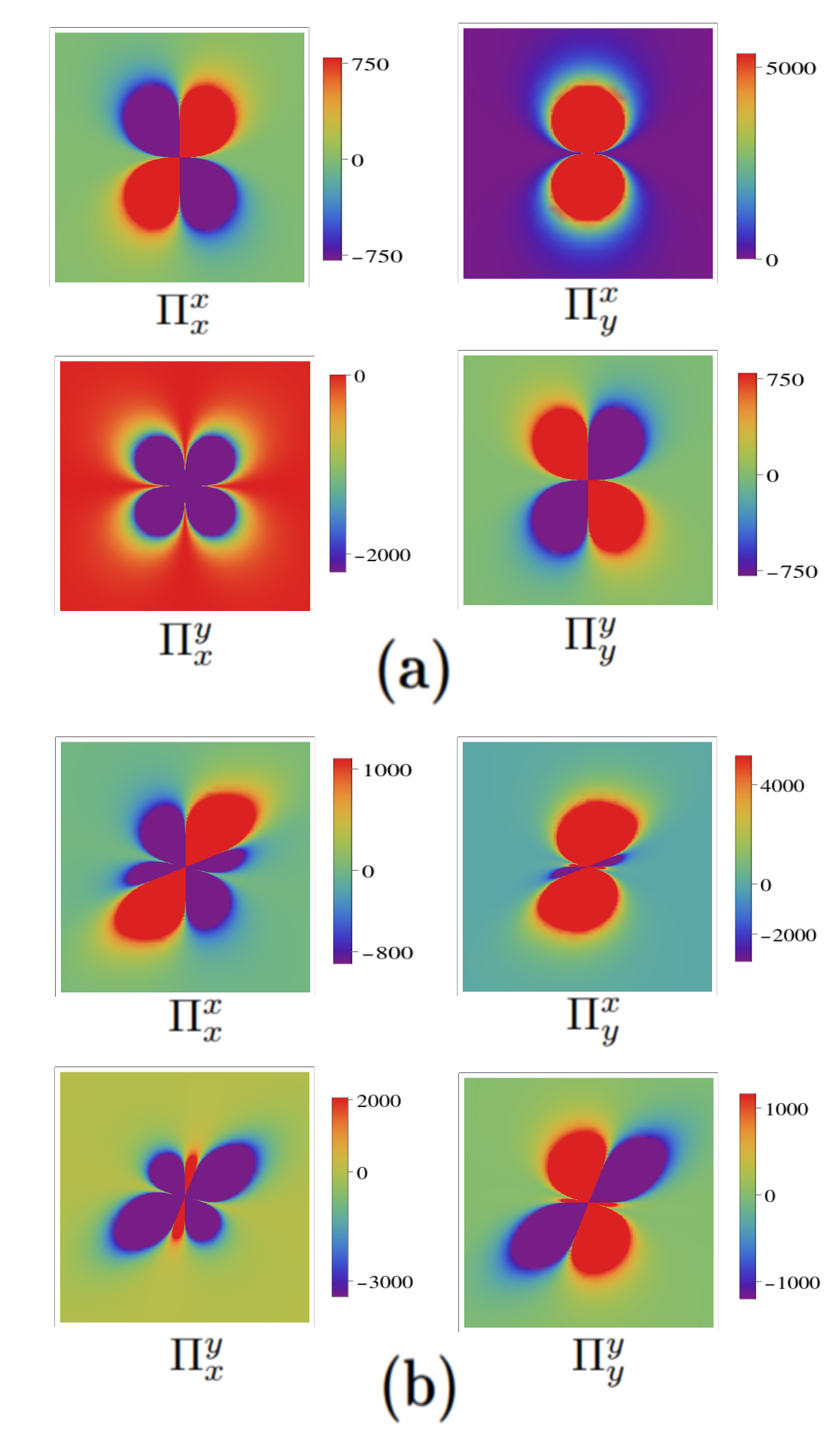}
    \captionsetup{format=plain, font=small, labelfont=bf, justification = raggedright}
    \caption{Density plots of the SBCP components, in units of $e^2\tilde S_0/(16\alpha k_h^6)$, for the $\varepsilon_+({\bf k})$ band in the $k_x$-$k_y$ plane: (a) pure Rashba case ($\beta =0 $); 
    (b) Rashba-Dresselhaus case ($\beta/\alpha =0.4$).}
    \label{fig:sbcp}
\end{figure}

In Fig. (\ref{fig:sbcp}) the density plots of the SBCP components are shown. 
They display a multipolar nature. 
In the absence of the Dresselhaus interaction (Fig. (\ref{fig:sbcp}(a)) all four SBCP components are even functions in the $k$-space. 
Further, $\Gamma^x_y$ and $\Gamma^y_x$ are symmetric under reflection about the $k_x$ and $k_y$ axes, whereas $\Gamma^x_x$ and $\Gamma^y_y$ are anti-symmetric under such reflections.
This property of the SBCP components can be in found in 2D electron gases as well \cite{zhang2024intrinsic}.
The inclusion of the Dresselhaus term distorts the symmetric distributions as shown in Fig. (\ref{fig:sbcp}(b)). 
 
\begin{figure}
\centering
\includegraphics[width=\linewidth]{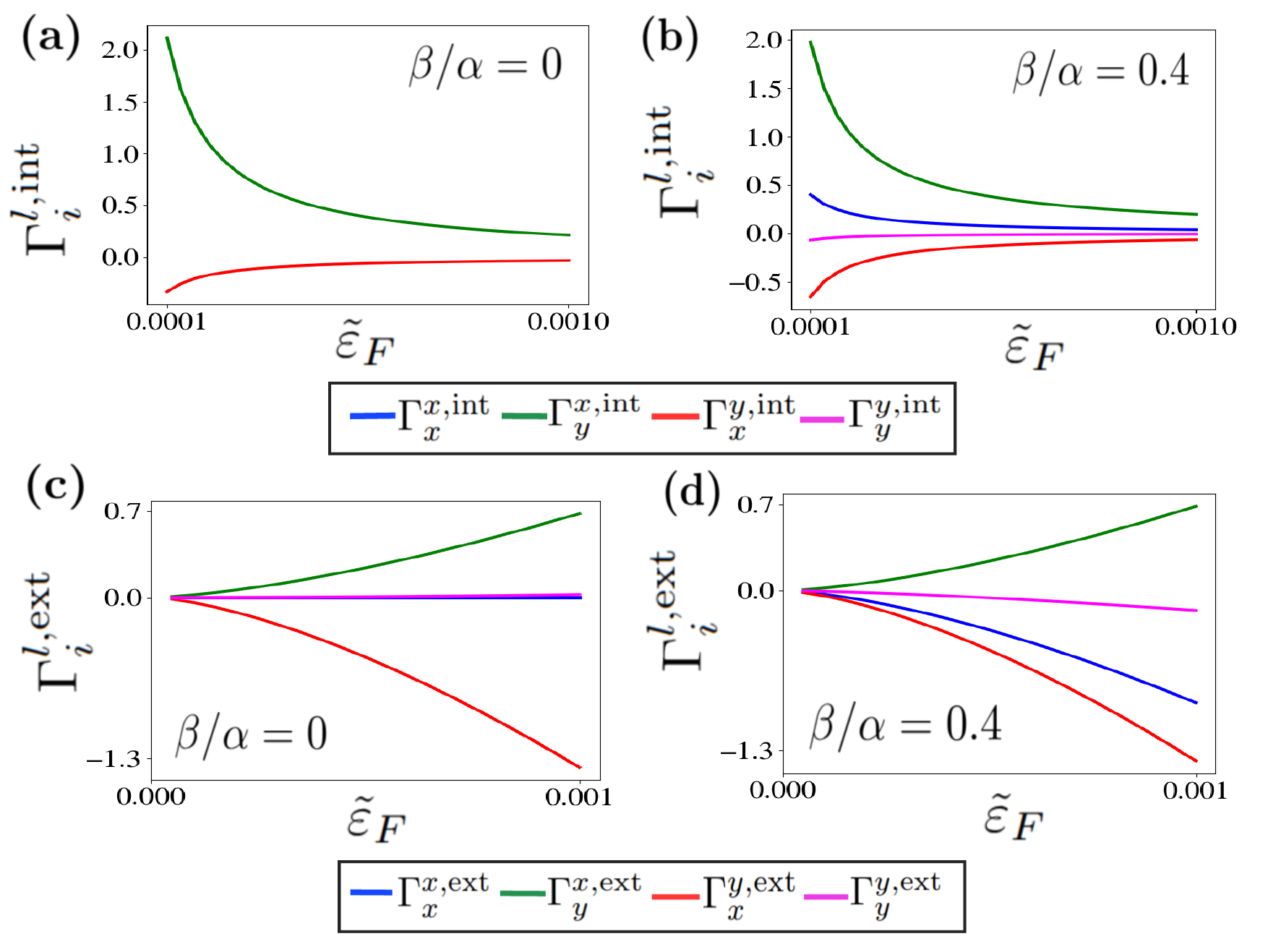}
\captionsetup{format=plain, font=small, labelfont=bf, justification = raggedright}
\caption{Second-order spin conductivity vs. Fermi energy (in units of $\varepsilon_h$). {\bf Top panel}: variation of intrinsic second-order spin conductivity with Fermi energy for (a) $\beta/\alpha=0$ (b) $\beta/\alpha=0.4$. The scale of $\Gamma^{l,\text{int}}_i$ is set at $\Gamma_0\times10^6$, with 
$\Gamma_0 = e^2\tilde S_0/(16\alpha k_h^4)$.
{\bf Bottom panel}: extrinsic contribution to second-order spin conductivity (in units of $\Gamma_0$) for (c) $\beta/\alpha=0$ (d) $\beta/\alpha=0.4$. The relaxation time $\tau$ is chosen to be $1\:\rm{ps}$.}
\label{fig:spin conductivity}
\end{figure}

The intrinsic second-order in-plane spin conductivity is derived from the SBCP as shown in Eq. (\ref{2nd-order-spin conductivity}). 
They are plotted as functions of $\tilde{\varepsilon}_F$ in Fig. (\ref{fig:spin conductivity}) along with the corresponding extrinsic components.
When both the Rashba and Dresselhaus terms are present, all the four components of the second-order 
spin conductivity are non-zero. In the pure Rashba case, the collinearly polarized spin currents 
($\Gamma^x_x$ and $\Gamma^y_y$) vanish, while other components ($\Gamma^x_y$ and $\Gamma^y_x$)  survive. 
For 2D electron gases with $k$-linear Rashba interaction, the absence of 
$\Gamma^x_x$ and $\Gamma^y_y$ is attributed to the $\pi/2$ angle-locking between the 
spin polarization and the wave-vector ${\bf k}$ \cite{zhang2024intrinsic}. 
This is also the case for the $k$-cubic hole gas. It has been shown in the contour plots (Fig. (\ref{fig:contour})) given in Appendix (\ref{App:contour}) that the spin polarization lies tangential to the Fermi contour in the pure Rashba case. 

The vanishing of $\Gamma_x^x$ and $\Gamma^y_y$ can also be understood from the mirror anti-symmetry of $\Pi^x_x$ and $\Pi^y_y$, as shown in Fig. (\ref{fig:sbcp}(a)), which integrate to zero over the {\bf k}-spcae.
The introduction of the Dresselhaus term makes the band structure anisotropic and removes the mirror anti-symmetry, allowing $\Gamma^x_x$ and $\Gamma^y_y$ to become finite as well. 
Thus, the band anisotropy due to Rashba-Dresselhaus spin-orbit coupling in 2DHG 
leads to the generation of all $\mathcal{T}$-allowed in-plane spin currents.
Moreover, we obtain both transverse and longitudinal spin currents in the second-order, irrespective of whether the band structure is anisotropic or not, unlike the first-order where the longitudinal spin current arises only in the anisotropic case. 

The extrinsic spin conductivity (see Appendix \ref{App:extrinsic}) is contributed by the $\tau^2$-dependent term in Eq. (\ref{2nd-order-spin conductivity}) 
which is symmetric under $\mathcal{T}$. 
 
Unlike the electron gas \cite{zhang2024intrinsic} the extrinsic spin conductivity for the hole gas is found to  increase in magnitude with the Fermi energy. 
However, the magnitudes of the extrinsic components are negligibly small compared to the intrinsic ones. As shown in Fig. (\ref{fig:spin conductivity}), $\Gamma^{l,\rm ext}_i/\Gamma^{l,\rm int}_i\sim 10^{-6}$. 
Converting to realistic units, we get $\Gamma_0 = 0.625\: \rm e\:\mu m/V$, which is of the same order of magnitude as the second-order spin conductivity, reported in electron gas\cite{zhang2024intrinsic}.
Hence, the extrinsic second-order spin conductivity in electron and hole gases are of comparable magnitude while the intrinsic second-order spin conductivity is enhanced in the heavy hole gas. This enhancement of the intrinsic second-order spin conductivity is due to the small spin-split band gap and the $k$-cubic nature of the bands.
Using the definition of $k_h$ and the parameter values given in section(\ref{secIII}), $k_h$ is estimated to be around $1\: \rm nm^{-1}$. Consequently, the energy scale $\varepsilon_h$ turns out to be $0.2\: \rm eV$. The range of Fermi energy, in figures (\ref{fig:linear_spin conductivity}) and (\ref{fig:spin conductivity}), in meV is $0.02-0.2$ meV. 
Unlike the 2D electron gas, here the local maxima of the lower energy branch puts a cut-off on the momentum scale and restricts the energy range.  

\section{Two-dimensional hole gas with Rashba SOC subjected to linearly polarized 
electromagnetic radiation}\label{secIV}
As an alternative method of controlling the band anisotropy of the system, 
we irradiate linearly polarized electromagnetic radiation over the 2DHG with Rashba spin-orbit coupling.
We consider a linearly polarized radiation with frequency $\omega $, propagating perpendicular to the system.
Assuming the wavelength of radiation to be much larger than the lattice constant, the radiation field can be described by the vector potential, ${\bf A}(t)= A_0[\sin(\omega t),\sin(\omega t+\phi_0)]$, with amplitude $A_0$ and phase difference $\phi_0$ between the two spatial components\cite{Firoz}. For linear polarization, $\phi_0$ is chosen to be zero. 

\begin{figure}[htbp]
    \centering
    \includegraphics[trim=0.5cm 0cm 0cm 0cm, clip, width=1\linewidth]{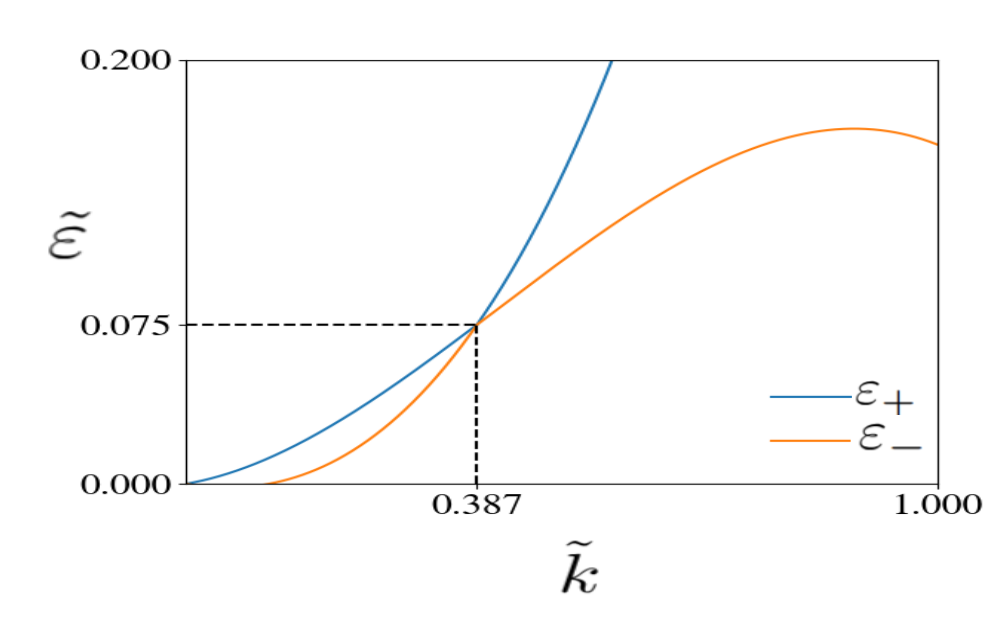}
    \captionsetup{format=plain, font=small, labelfont=bf, justification = raggedright}
    \caption{Dispersion of Rashba spin-orbit coupled hole gas in the presence of 
    electromagnetic radiation for $\phi=\pi/4$, $\tilde A= 0.15$ and  $\alpha_0=0.04$. 
    The band degeneracy occurs at $\tilde{k}=0.387$ and the corresponding energy is about 
    0.075 in the given scale.}
    \label{fig:disp2}
\end{figure}

Using the Floquet-Magnus expansion \cite{floquet-magnus} and taking the high-frequency limit, 
the final Hamiltonian becomes \cite{Firoz}
\begin{equation}\label{em-Hamiltonian}
    H = \frac{\hbar^2k^2}{2m} \mathbb{I} + i\alpha(k_{-}^{3}\sigma_{+} - k_{+}^{3}\sigma_{-}) - \alpha A {\bs \sigma} \cdot {\bf k}, 
\end{equation}
where $ A = 3(eA_{0}/\hbar)^2$. The additional term, $\alpha A {\bs \sigma} \cdot {\bf k}$ which imparts anisotropy to the band dispersion, is a frequency-independent zeroth-order correction in the Floquet-Magnus expansion. The first-order correction vanishes due to linear polarization of the radiation, while the higher-order frequency dependent terms have been ignored\cite{Firoz}.
Eq. (\ref{em-Hamiltonian}) is taken as the unperturbed Hamiltonian, over which a static electric field is added as the perturbation, leading to a spin Hall response. 
The energy dispersion is given by
\begin{equation}
    \varepsilon(k) = \frac{\hbar^2k^2}{2m} \pm \alpha k \sqrt{(k^2 - A\sin(2\phi))^2 + A^2\cos^{2}(2\phi)}
\end{equation}
with $\tan\phi= k_{y}/k_{x}$.
The two dispersion bands are shown in Fig. (\ref{fig:disp2}).
The eigen spinors are,
\begin{equation}
    |\pm\rangle = \frac{1}{\sqrt{2}}\begin{pmatrix}1 \\ \pm e^{i\psi} \end{pmatrix},
\end{equation}
where  $\tan\psi = C/D $, with 
$C = A\sin\phi+  k^2 \cos(3\phi) $ and $D = A\cos\phi- k^2 \sin(3\phi)$.
The energy band gap is given by 
\begin{equation}
    \varepsilon_g = 2\alpha k \sqrt{(k^2 - A\sin(2\phi))^2 + A^2\cos^{2}(2\phi)}.
\end{equation}
There are degeneracy points at $k=\sqrt{A}$ with
$\phi=(2s+1)\pi/4$ with $s=0,1,2,3$. 
We introduce dimensionless variables by scaling the $k$ by a typical wave-vector $k_0$ ($\tilde{k}=k/k_0$) and the energy is scaled by $\gamma= \hbar^2k_0^2/(2m)$ ($\varepsilon/\gamma$). 
Consequently, the other parameters get scaled as, $\tilde{A} = A/k_0^2 $, $\alpha_0 = \alpha k_0^3/\gamma$, $\tilde S_0 = S_0k_0$, and $\tilde S_1 = S_1k_0^2$.

The validity of this analysis depends, critically on the high-frequency limit which is explicitly given by, $\hbar\omega\gg(\gamma , \alpha k_0^3, eA_0 \hbar k_0 /m, 3eA_0 \alpha k_0^2/\hbar, e^2 A^2_0\alpha k_0/\hbar^2 )$\cite{Firoz}. Further, the off-resonant condition states that $\hbar\omega\gg \varepsilon_g$. Following Ref. \cite{Firoz} the photon energy should be chosen around $\hbar\omega=0.2-0.3\: \rm eV$, which is much larger, not only compared to the spin-split gap of the heavy hole branch, but also the heavy hole-light hole splitting\cite{Effective_spins,murakami2003dissipationless}. 
By staying in the high-frequency and off-resonant regime, effects like absorption and photovoltaic charge and spin currents can be suppressed.

\subsection{Spin Berry curvature and linear spin conductivity}
\begin{figure}[htbp]
    \centering
    \includegraphics[trim=0cm 0cm 0cm 0cm, clip, width=1\linewidth]{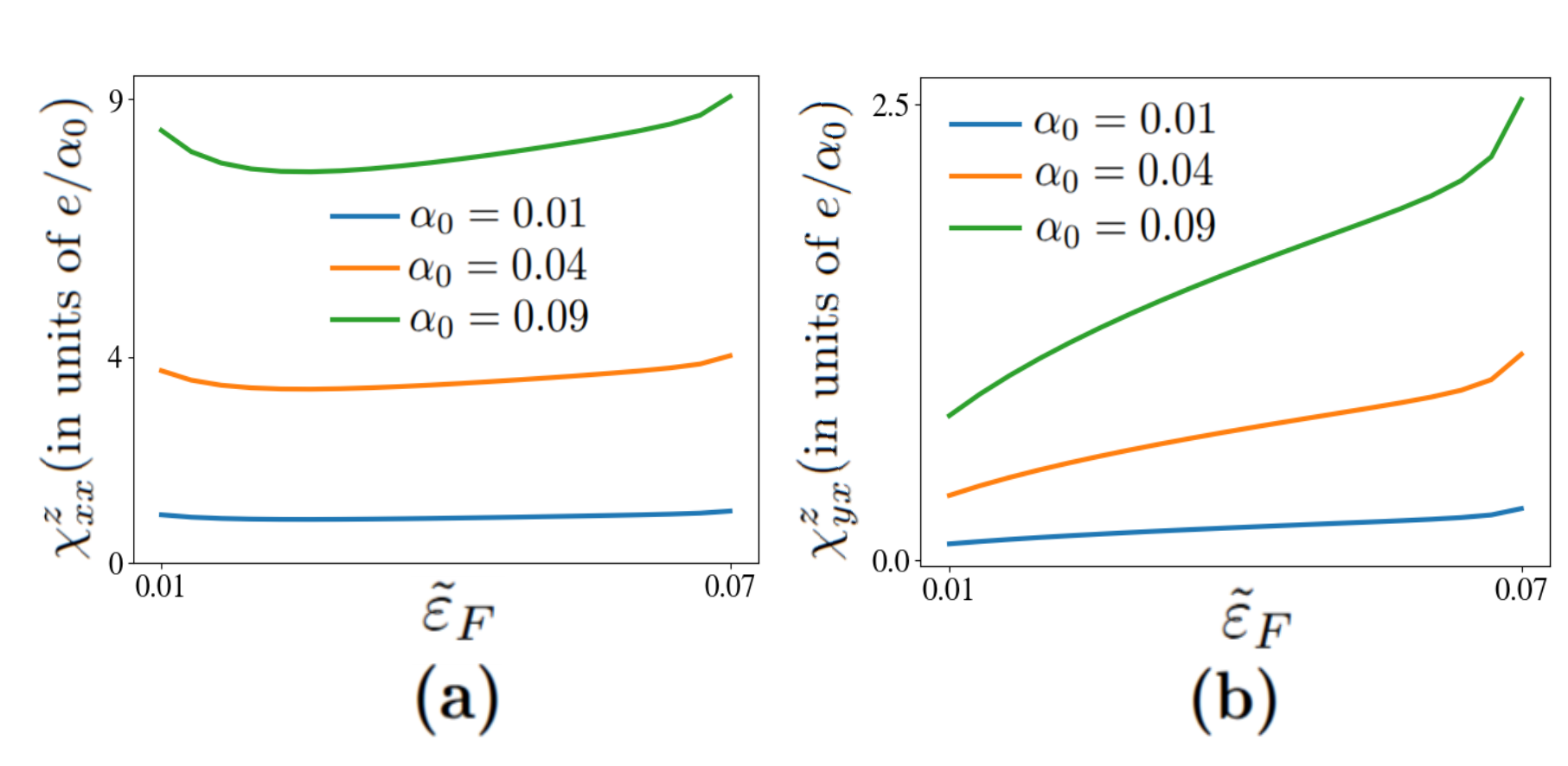}
    \captionsetup{format=plain, font=small, labelfont=bf, justification = raggedright}
    \caption{({\bf a}) Linear longitudinal spin conductivity and 
    ({\bf b}) linear spin Hall conductivity (in units of $e/\alpha_0$) as a function of 
    the scaled Fermi energy ($\varepsilon/\gamma$) for different values of $\alpha$. 
    }
    \label{fig:em_linear}
\end{figure}

It is already known that only the out-of-plane components of the SBC tensor 
will survive. Since, the electric field is along the $x$ direction, 
the spin Hall current is computed along the $y$ direction. 
The transverse components of the SBC turns out to be,
\be 
\Omega^z_{yx,\pm} = \mp 
\frac{3\sin\phi}{2 \alpha_0 k_0^2 \tilde{k}}
\l[ \frac{ 3\tilde{k}^2 \cos(2\phi - \psi) + \tilde{A}\cos\psi} 
{\tilde{k^4} + \tilde{A}^2 - 2 \tilde A \tilde k^2 \sin(2\phi)} \r].
\ee
while its longitudinal component is given by
\be 
\Omega^z_{xx,\pm} = \mp 
\frac{3\cos\phi}{2 \alpha_0 k_0^2 \tilde{k}}
\l[ \frac{ 3\tilde{k}^2 \cos(2\phi - \psi) + \tilde{A}\cos\psi} 
{\tilde{k^4} + \tilde{A}^2 - 2 \tilde A \tilde k^2 \sin(2\phi)} \r].
\ee
The linear spin Hall conductivity $ \chi^z_{yx}$ and the linear longitudinal spin conductivity $\chi^z_{xx}$ are plotted in figures (\ref{fig:em_linear}(b)) and (\ref{fig:em_linear}(a)) respectively, as functions of Fermi energy for different values 
of $\alpha_0 $.

The spin Hall resonance occurs as the Fermi energy approaches the degeneracy. This has been previously studied in 2D electron and hole gases in the presence of external magnetic field \cite{SHR1,SHR2,SHR3}.

\subsection{Spin Berry curvature polarizability and second-order spin conductivity}
Following the analysis done in sections (\ref{sec:II}) and (\ref{secIII}), 
the SBCP components for the in-plane spin current have been calculated using the effective spin operators (Eqs. (\ref{sx_modified}-\ref{sz_modified})).

\begin{figure}
    \centering
    \includegraphics[trim=0cm 0cm 0cm 0cm, clip, width=1\linewidth]{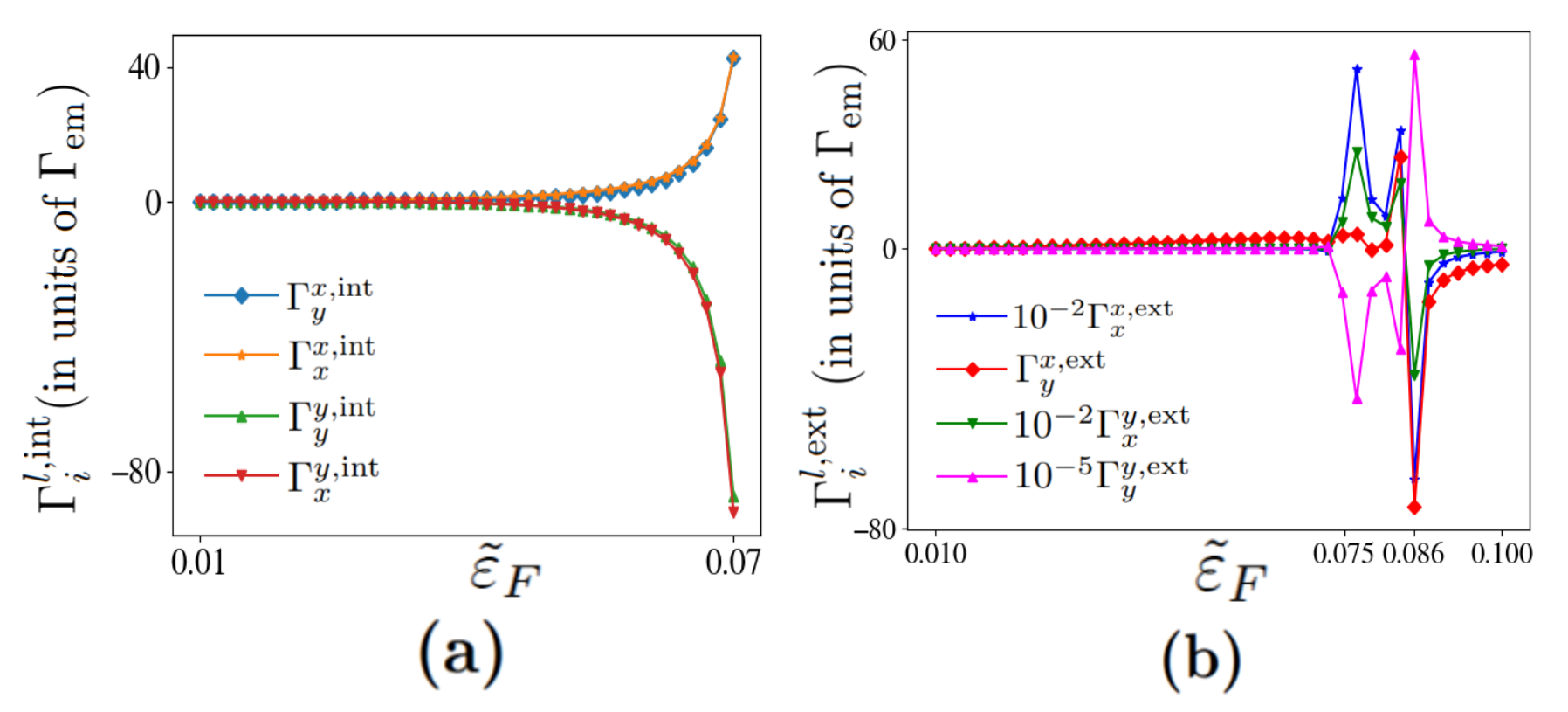}
    \captionsetup{format=plain, font=small, labelfont=bf, justification = raggedright}
    \caption{{\bf (a)} Second-order intrinsic spin conductivity and {\bf (b)} Second-order extrinsic spin conductivity (in units of $\Gamma_{\rm{em}}=e^2\alpha_0^2\tilde S_0/(\gamma k_0) $ as a function of the scaled Fermi energy. Here $\tau = 1 $ ps.}
    \label{fig:em_spin conductivity}
\end{figure}

The plots of the second-order intrinsic spin conductivity versus the Fermi energy are shown in 
Fig. (\ref{fig:em_spin conductivity}(a)). The components appear to diverge as the energy approaches 
the resonance point, similar to the linear spin conductivity.
The extrinsic contribution appears in the second-order spin conductivity from the non-linear Drude term in 
Eq. (\ref{2nd-order-spin conductivity}). 
The four in-plane components of the extrinsic spin conductivity are plotted against the Fermi energy in Fig. (\ref{fig:em_spin conductivity}(b)). However, unlike section (\ref{secIII}) here the extrinsic conductivities have varying magnitudes which are greater than those of the intrinsic conductivities. Fig. (\ref{fig:em_spin conductivity}(b)) also shows that these components have different peak values at resonance. They can have both positive and negative peaks, and the peaks obtained for the extrinsic case are slightly shifted from those obtained for the intrinsic one. 
While the band geometry-induced intrinsic spin response diverges at resonance point where the band gap vanishes, the extrinsic spin conductivity exhibits finite peaks around the resonance point. This shows that due to disorder broadening, the observed resonance peaks will be finite and broad.
Heating of the sample may also contribute to this broadening as shown in Ref. \cite{Firoz}.
For a more detailed analysis of the spin responses at the resonance point and their scaling with Fermi energy away from it, an effective Hamiltonian around the radiation-induced degeneracies may be required. This is beyond the scope of the present article, but is an interesting future prospect.

In actual experiments, even if photovoltaic spin or rectification currents appear, they can be distinguished from the spin Hall current, generated by the dc field. The background spin current due to photovoltaic effects can be estimated by turning the dc electric field off and should be subtracted from the actual results. 
Further, a distinctive feature of the spin Hall current is their resonance, when the Fermi energy crosses the radiation-induced band degeneracy, as discussed above.

\begin{table}
\begin{tabular}{|c|c|c|c|c|}
\hline
{} & \multicolumn{4}{c|}{In-plane}\\ 
\cline{1-5}
 {\bf First-order} & \:\:\:$\chi^x_x$\:\:\:\: & $\chi^y_x$ & \:\:$\chi^x_y$\:\:\:\: & \:\:$\chi^y_y$\:\: \\ 
\hline
Isotropic & $\times$ & $\times$ & $\times$ & $\times$\\ 
\cline{1-5}
Anisotropic & $\times$ & $\times$ & $\times$ & $\times$\\ 
\hline
\hline
{\bf Second-order} & $\Gamma^x_x$ & $\Gamma^y_x$ & $\Gamma^x_y$ & $\Gamma^y_y$\\ 
\hline
Isotropic & $\times$ & $\checkmark$ & $\checkmark$ & $\times$\\ 
\cline{1-5}
Anisotropic & $\checkmark$ & $\checkmark$ & $\checkmark$ & $\checkmark$\\ 
\hline
{} & \multicolumn{2}{|c|}{Longitudinal} & \multicolumn{ 2}{|c|}{Transverse}\\ 
\hline
\end{tabular}
\caption{All the vanishing and non-vanishing components of the first- and second-order spin conductivity tensor are depicted as  $\checkmark$:\: Present, $\times$:\: Absent}
\label{tab}
\end{table}

\section{Summary and Discussion}
In this work, we have shown that 
the first non-vanishing in-plane spin current is non-linear (second-order),
whereas the linear spin current is out-of-plane
for any generic time-reversal symmetric 2D system.
A summary of the vanishing and non-vanishing components of the linear and non-linear spin conductivity tensors are presented in Table (\ref{tab}).
As a specific example, we considered two-dimensional heavy hole gas 
with spin-orbit coupling formed at the III-V semiconductor heterojunctions.
It is shown that the linear spin Hall conductivity for pure
Dresselhaus coupling ($\alpha = 0, \beta \neq 0$) is reduced by a factor of 2/3, 
as compared to the pure Rashba interaction ($\alpha \neq 0, \beta = 0$). The anisotropic Rashba-Dresselhaus spin-orbit coupling also induces a non-zero longitudinal spin conductivity, which otherwise vanishes in the pure Rashba or pure Dresselhaus case. It is also interesting to note that the second-order spin current can have a longitudinal component even when the spin-orbit interaction is isotropic.

For the pure Rashba case, the second-order in-plane spin current is always normal to the 
spin-polarization. The band anisotropy, introduced by either the Dresselhaus spin-orbit interaction
or an external electromagnetic field, generates additional second-order collinearly polarized spin currents, which are 
parallel to the spin-polarization.
Therefore by tuning the ratio of the Rashba and Dresselhaus coupling strengths \cite{Gate_control_soc} or 
the amplitude of the electromagnetic radiation, the multiple in-plane spin currents 
can be controlled. 
The anisotropy may be used as a switch to turn on or off certain 
second-order spin currents that vanish in the pure Rashba limit. 
 
It is also observed that the extrinsic non-linear spin current is vanishingly small, compared to the intrinsic one, for the $k$-cubic hole gas with Rashba-Dresselhaus spin-orbit coupling, which is in contrast to the $k$-linear 2D electron gas.
However, in the presence of a linearly polarized electromagnetic radiation, the hole gas can host extrinsic non-linear spin currents equalling or exceeding the intrinsic ones in magnitude.
The electromagnetic radiation also causes spin Hall resonance at a certain Fermi energy 
which can lead to a giant non-linear spin current.
As discussed in section (\ref{secIV}), the different second-order spin currents can be distinguished based on their behaviour around the resonance point.

The calculated non-linear bulk spin currents can be probed in various ways. These include magneto-optic Kerr effect (MOKE) to detect spin accumulation at the sample edges\cite{Kato-first-paper,SHE-expt-semiconductor,SHE-expt-heavy-metal}, spin to charge conversion using the inverse spin Hall effect\cite{Saitoh} and detecting the second harmonic response of the applied ac electric field\cite{2nd-harmonic-spin-current,article_B}, and the detection of spin-orbit torque exerted by the spin currents when injected into a magnetic material\cite{ST-FMR, spin-torque-switching,Garello_magnetization-switching}. 
Among these, MOKE and spin-orbit torque experiments are most relevant to this work since the focus is on dc spin conductivity. Both of these techniques are capable of resolving the in-plane and out-of-plane spin polarization. 
Particularly, in our case where the first-order spin current is entirely out-of-plane and the second-order spin current lies in-plane, detecting an in-plane spin polarization would provide a characteristic signature of the second-order response.
Having said that, one must also note that the in-plane spins in hole gases are generally subjected to stronger precession and relaxation\cite{hole-spin-relaxation}, which may hinder their detection at the sample edges. However, the in-plane spin relaxation lifetimes are anisotropic\cite{Intro_1} and can be controlled experimentally\cite{spin-relaxation-control}.
Spin diffusion length is another important parameter which is required to predict the spin accumulation profile, and to choose the appropriate sample width. Using typical parameter values for semiconductor systems\cite{SHE-review}, the spin diffusion length can be estimated to be $0.1-1\: \rm \mu m$, which sets the width of the Hall bars at a few microns\cite{SHE_expt}.  

The collinearly polarized spin currents studied in this article are particularly important for field-free magnetization switching \cite{Hui_Wang} in spin-orbit torque (SOT)-driven spintronic memory devices\cite{SOT-MRAM}.  
Furthermore, non-linear spin currents, similar to their charge counterparts, may also be used to distinguish between $k$-linear and $k$-cubic spin-orbit interactions\cite{article_C}, as already mentioned in section(\ref{secIII}). 
Exploring possible non-linear Shubkinov-de Haas oscillations\cite{article_E} in these non-linear spin conductivities is another interesting future direction, which may be used to probe Fermi-surface spin textures and the underlying spin-orbit coupling mechanisms. 
\\

{\bf AUTHOR DECLARATION}\\

The authors have no conflicts to disclose.\\

{\bf ACKNOWLEDGEMENT}\\

SC acknowledges IIT Kanpur for providing Ph.D. fellowship.\\

{\bf DATA AVAILABILITY STATEMENT}\\

The data that support the findings of this study are available from the corresponding author upon reasonable request.


\appendix
\section{Derivation of SBC and SBCP components for a two-level system}\label{App:2nd-order}
In this appendix, we derive the electric field induced first and second-order spin current.
The normalized perturbed eigenstate up to the second-order is given by \cite{Sakurai, bransdenquantum}
\begin{equation}
    \vert \tilde n \rangle = \vert n^{(0)} \rangle + \vert n^{(1)} \rangle + \vert n ^{(2)} \rangle,
\end{equation}
where
\be \label{1st-order}
\vert n^{(1)} \ra = \sum_{m \neq n } \frac{- e {\bf E} \cdot {\bf A}_{mn} }
{\varepsilon_m^{(0)} - \varepsilon_n^{(0)} }
\vert m^{(0)}\ra,
\ee
and
\begin{widetext}
\begin{align} \label{2nd-order}
 |n^{(2)}\rangle = e^2\left[\sum_{m\neq n} \left[ \sum_{p\neq n}\frac{({\bf E\cdot A}_{mp})({\bf E\cdot A}_{pn})}{\left(\varepsilon^{(0)}_n-\varepsilon^{(0)}_p \right)\left(\varepsilon^{(0)}_n-\varepsilon^{(0)}_m \right)} 
 - \frac{({\bf E\cdot A}_{mn})({\bf E\cdot A}_{nn})}{\left(\varepsilon^{(0)}_n-\varepsilon^{(0)}_m \right)^2}\right] |m^{(0)}\rangle 
  -\frac{1}{2}\sum_{p \neq n} \frac{\vert({\bf E\cdot A}_{np})\vert^2}{\left(\varepsilon^{(0)}_n-\varepsilon^{(0)}_p \right)^2} \vert n^{(0)}\rangle \right].
\end{align}
\end{widetext}
Since we are working with a two-level system, for a given state $|n^{(0)}\rangle$ , both indices $m$ and $p$ denote the same state $(m=p \neq n)$. Hence, the summations in Eq. (\ref{1st-order}) and (\ref{2nd-order}) are not required. Let us also denote $(\varepsilon^{(0)}_n-\varepsilon^{(0)}_m) = (\varepsilon^{(0)}_n-\varepsilon^{(0)}_p)$ as $\varepsilon_g$, the energy gap between the two states. Therefore, the first-order and second-order corrections to the state are simplified to
\begin{equation}
    \vert n^{(1)} \ra =-\frac{ e}{\varepsilon_g} {\bf E} \cdot {\bf A}_{mn} \vert m^{(0)}\ra,
\end{equation}
and
\begin{widetext}
\begin{equation}
    |n^{(2)}\rangle = \frac{e^2}{\varepsilon_g^2} \left[\left(({\bf E\cdot A}_{mm})({\bf E\cdot A}_{mn}) - ({\bf E\cdot A}_{mn})({\bf E\cdot A}_{nn}) \right)|m^{(0)}\rangle 
     - \frac{1}{2} \vert {\bf E\cdot A}_{nm} \vert^2 \vert n^{(0)}\rangle  \right],
\end{equation}
\end{widetext}
respectively with $m \neq n$. 
Now, the expectation value of the spin current operator in the modified state is given by $\langle\tilde{n}|\hat{j}^l_{i}|\tilde{n}\rangle$.
Collecting the terms linear in $E$ we get,
\begin{equation}
    \tilde{j}^{l,(1)}_{i,n}  = \langle n^{(0)}|\hat{j}^l_{i}|n^{(1)}\rangle + \langle n^{(1)}|\hat{j}^l_{i}|n^{(0)}\rangle,
\end{equation}
which gives
\begin{equation}\label{j1}
  \tilde{j}^{l,(1)}_{i,n} = \frac{2e}{\varepsilon_g}\text{Re} \left\{\langle n^{(0)}|\hat{j}^l_{i}|m^{(0)}\rangle A^j_{mn} \right\}E_j.
\end{equation}
The inter-band Berry connection can be written in terms of the group velocity operator $\hat{\bf v} = \frac{1}{\hbar}\nabla_{\bf k}\hat{H}$ as, 
\begin{equation}\label{berry-connection}
    {\bf A}_{mn} = \frac{i}{\varepsilon_g}\langle m^{(0)}|\hat{\bf v}|n^{(0)}\rangle.
\end{equation}
Putting Eq. (\ref{berry-connection}) in Eq. (\ref{j1}) and following Eq. (\ref{modified-spin-current}) we obtain the expression of SBC as
\begin{equation}
    \Omega^l_{ij,n} = \frac{2\hbar}{\varepsilon_g^2}\ \Im\left[\langle n^{(0)} |\hat{j}^{l}_{i}|m^{(0)}\rangle\langle m^{(0)}|\hat{v}_{j}|n^{(0)}\rangle\right].
\end{equation}
Denoting the eigenstates as $\vert \pm \rangle$, the expression is rewritten in the following form
\begin{equation}
    \Omega^l_{ij,\pm} = \frac{2\hbar}{\varepsilon_g^2}\ \Im{\Big[\langle \pm |\hat{j}^{l}_{i}|\mp\rangle\langle \mp|\hat{v}_{j}|\pm\rangle}\Big],
\end{equation}
as given in section(\ref{sec:IIB}).
Similarly, collecting the terms quadratic in $E$, we get
\begin{equation} \label{j2_1}
    \tilde{j}^{l,(2)}_{i,n}  = \langle n^{(1)}|\hat{j}^l_{i}|n^{(1)}\rangle + \langle n^{(0)}|\hat{j}^l_{i}|n^{(2)}\rangle + \langle n^{(2)}|\hat{j}^l_{i}|n^{(0)}\rangle. 
\end{equation}
Out of this, the terms that contribute to in-plane second-order spin current can be written as


\begin{equation}
    \tilde j^{x/y,(2)}_{i,n} = \frac{e^2}{\varepsilon_g^2} \vert {\bf E}\cdot{\bf A}_{nm}\vert^2 \left(\langle m^{(0)}\vert \hat j_{i}\vert m^{(0)}\rangle - \langle n^{(0)}\vert \hat j_{i}\vert n^{(0)}\rangle \right).
\end{equation}
Writing ${\bf A}_{nm}$ in terms of the velocity operators (Eq. (\ref{berry-connection})) and denoting the unperturbed states by $\vert \pm\rangle$ we obtain,
\begin{align}
   \tilde j^{x/y,(2)}_{i,n} = \frac{e^2\hbar^2}{\varepsilon_g^4}  \langle \mp|\hat v_{j}|\pm\rangle &\langle \pm|\hat v_{k}|\mp\rangle \notag\\
   &\left(\langle \mp|\hat{j}^{l}_{i}\vert\mp\rangle -\langle \pm|\hat{j}^{l}_{i}|\pm\rangle\right)E_jE_k.
\end{align}
This leads to the expression for SBCP given in Eq. (\ref{sbcp2}) in section(\ref{sec:IIB}), following Eq. (\ref{modified-spin-current}). \\


Eq.(\ref{j2_1}) gives another term but it does not contribute to the in-plane spin current,
\begin{equation}\label{j2_3}
  \tilde{\mathcal{J}}^{l,(2)}_{i,n} = \frac{e^2}{\varepsilon_g^2}\left[2\text{Re} \left\{\langle n^{(0)}|\hat{j}^l_{i}|m^{(0)}\rangle A^j_{mn} \right\} (A^k_{mm}-A^k_{nn})E_jE_k\right].
\end{equation}
The factor $2\text{Re}\left\{\langle n^{(0)}|\hat{j}^l_{i}|m^{(0)}\rangle A^j_{mn}\right\}$ vanishes for $l=x,y$, as explained in section(\ref{sec:IIB}).  
Also, we note that there are two gauge dependent terms ($A_{mm}$ and $A_{nn}$) in Eq. (\ref{j2_3}). 
If there are more than one gauge dependent terms present in a given expression, 
one must use all the eigenvectors in the same gauge to calculate them. 
The eigenstates (Eq. (\ref{eigen})) used in section(\ref{sec:IIB}) are in the same gauge since the transformation of the phase $\psi\rightarrow\psi+\pi$ takes $\vert + \rangle$ to $\vert -\rangle$,
and it can be shown that the Berry connections for the two bands are the same, i.e ${\bf A}_{mm}={\bf A}_{nn}$. 
Hence Eq. (\ref{j2_3}) makes no contribution to the second-order spin current.

\section{The expressions of the SBCP components}

Using Eq. (\ref{sbcp2}) the four SBCP components, calculated for the Rashba-Dresselhaus heavy hole gas are as follows:
\begin{widetext}

\begin{align}
    &\Pi_x^{x,\pm} = \mp\frac{4\Pi_0}{\tilde k^5}\left[\tilde S_1\cos\phi\cos\phi' - 3\tilde S_0\sin\phi\sin\phi' + 2\tilde S_0 r \sin\phi(2\cos\psi+\cos\phi')\right],\\
    &\Pi^x_{y,\pm} = \mp\frac{4\Pi_0}{\tilde k^5}\left[\tilde S_1\sin\phi\cos\phi' - 3\tilde S_0\sin\phi\cos\phi' + 2\tilde S_0 r \sin\phi(2\sin\psi+\sin\phi')\right],\\
    &\Pi^y_{x,\pm} = \pm\frac{4\Pi_0}{\tilde k^5}\left[\tilde S_1\cos\phi\sin\phi' - 3\tilde S_0\cos\phi\sin\phi' + 2\tilde S_0 r\cos\phi(2\cos\psi + \cos\phi')\right],\\
    &\Pi^y_{y,\pm} = \pm\frac{4\Pi_0}{\tilde k^5}\left[\tilde S_1 \sin\phi\sin\phi' - 3\tilde S_0\cos\phi\cos\phi' + 2\tilde S_0 r\cos\phi (2\sin\psi+\sin\phi')\right],
\end{align}
where $\Pi_0$ is given by
\begin{equation*}
    \Pi_0 = \frac{\tilde S_0}{16\alpha k_h^6}\frac{\left(3\cos\phi' - 2 r \left(\left(3 \cos^2\phi + \sin^2\phi\right)\sin\psi - \sin2\phi \cos\psi\right)\right)^2}{\left(1 + r^2 - 2r\sin2\phi\right)^2}.
\end{equation*}
Similarly, for the Rashba 2DHG in the presence of high-frequency electromagnetic radiation, the SBCP components are given by
\begin{align}
    &\Pi^x_{x,\pm} = 4\Pi_{\rm em}\left[\pm\tilde S_0\tilde k\tilde A\alpha_0\sin\phi\cos\psi \mp \tilde k^3\left(3\tilde S_0\alpha_0\sin\phi\sin\phi' + \tilde S_1\cos\phi\cos\phi'\right)\right],\\
    &\Pi^x_{y,\pm} = 4\Pi_{\rm em}\left[\pm\tilde S_0\tilde k\tilde A\alpha_0 \sin\phi\sin\psi \mp \tilde k^3\left(-3\tilde S_0\alpha_0\sin\phi\cos\phi' + \tilde S_1\sin\phi\cos\phi'\right)\right],\\
    &\Pi^y_{x,\pm} = 4\Pi_{\rm em}\left[\mp\tilde S_0\tilde k\tilde A\alpha_0\cos\phi\cos\psi \pm \tilde k^3\left(3\tilde S_0\alpha_0\cos\phi\sin\phi' + \tilde S_1\cos\phi\sin\phi'\right)\right],\\
    &\Pi^y_{y,\pm} = 4\Pi_{\rm em}\left[\mp\tilde S_0\tilde k\tilde A\alpha_0\cos\phi\sin\psi \mp \tilde k^3\left(3\tilde S_0\alpha_0\cos\phi\cos\phi' - \tilde S_1\sin\phi\sin\phi'\right)\right],
\end{align}
where $\Pi_{\rm em}$ is given by
\begin{equation*}
    \Pi_{\rm em} = \frac{\alpha_0^2\tilde S_0}{\gamma k_0^3}\frac{(\cos\phi' - \tilde A\sin\psi)^2}{\left(\tilde A^2 + \tilde k^4 - 2\tilde A\tilde k^2\sin2\phi \right)^2}.
\end{equation*}
Here $\phi'=2\phi-\psi$ and all other parameters are same as given in section (\ref{secIV}).
\end{widetext}


\section{Spin polarization vector on the Fermi contours}\label{App:contour}

The local spin polarization vector can be defined as 
${\bf P}_\pm({\bf k}) = \la \pm \vert {\bf \hat S} \vert \pm \ra $,
where the spin operator is given by ${\bf\hat  S} = \hbar(\hat S_x, \hat S_y)$.
For 2D heavy hole gas,  using the eigenspinors $(|\pm\ra)$ given in 
section (\ref{secIII}), we obtain in polar coordinates
\begin{align}
    &\la \pm \vert {\hat S_x} \vert \pm \ra = -S_0k\sin\phi \pm S_1k^2\cos\phi' \\
    &\la \pm \vert {\hat S_y} \vert \pm \ra = S_0k\cos\phi \mp S_1k^2\sin\phi'
\end{align}
where $\phi' = 2\phi-\psi$ as given in section(\ref{secIII}) and the explicit expressions of $\cos \phi^\prime $ and $\sin \phi^\prime$ are given by,
\begin{align}
    \cos\phi' = \frac{ \sin\phi - r\cos\phi}{\sqrt{1 + r^2 - 2r\sin(2\phi)} },\\
    \sin\phi' = \frac{\cos\phi - r\sin\phi}{\sqrt{1 + r^2 - 2r\sin(2\phi)}}.
\end{align}

\begin{figure}[htbp]
    \centering
    \includegraphics[trim=0cm 0cm 0cm 0cm, clip, width=\linewidth]{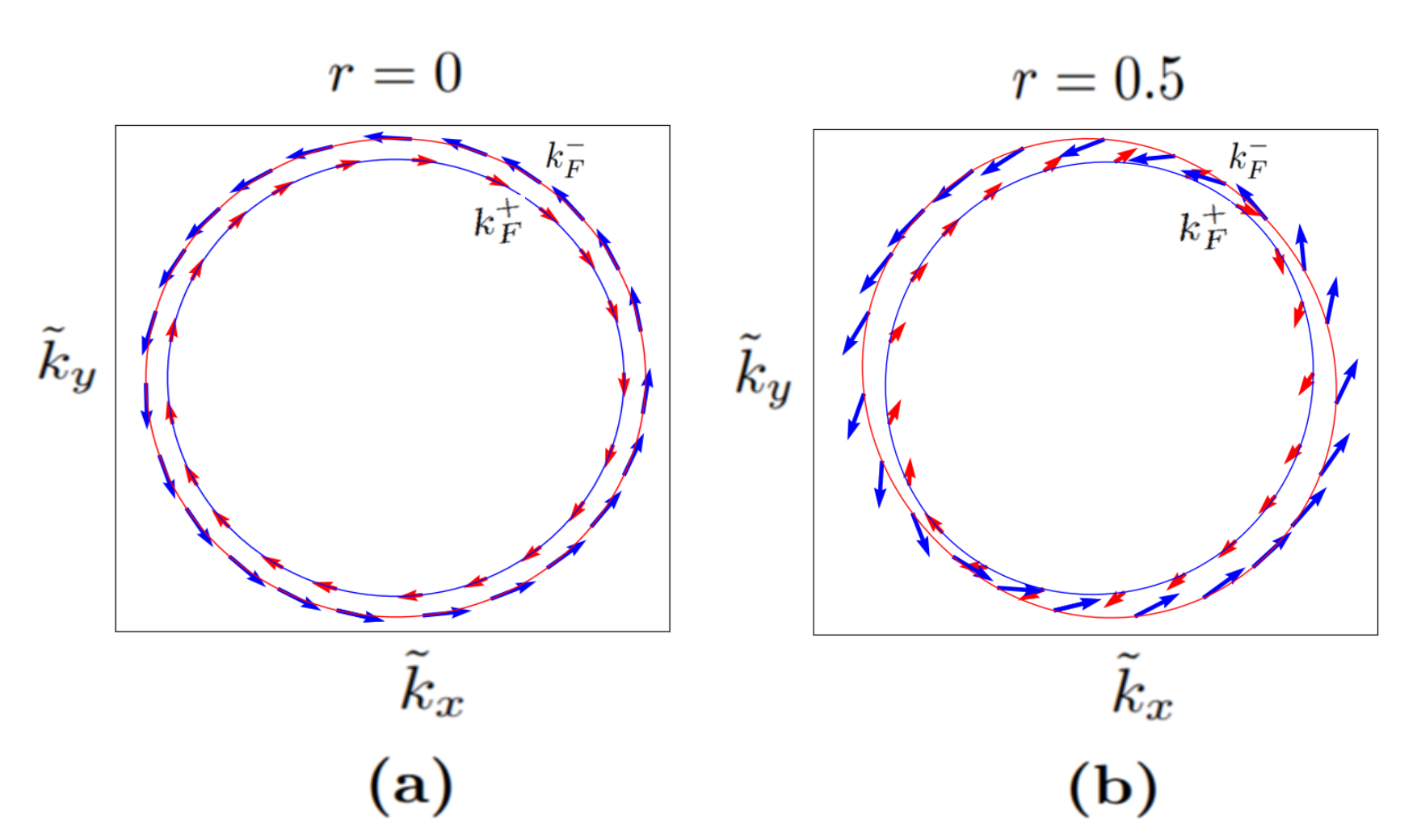}
    \captionsetup{format=plain, font=small, labelfont=bf, justification = raggedright}
    \caption{The local spin polarization vectors, ${\bf P}_{\pm}({\bf k})$ on the Fermi contours of the 
    2D hole gas for (a) pure Rashba ($r=0$) and (b) Rashba-Dresselhaus ($r = 0.5$) interactions.} 
    \label{fig:contour}
\end{figure}
The Fermi contours for the two bands and the local polarization vector ${\bf P}({\bf k}) $ 
are shown in Fig. (\ref{fig:contour}). 
Note that ${\bf P}({\bf k})$ is inversely proportional to the band gap $\varepsilon_g$. 
Therefore, the magnitude of $ {\bf P}({\bf k})$ vector is large near the points where
the band gap is small and it decreases with the increase of the band gap.


\section{Extrinsic spin current}\label{App:extrinsic}
The only extrinsic contribution to the spin  current comes from the non-linear Drude term given by
Eq. (\ref{2nd-order-spin conductivity}),
\begin{equation}
    \Gamma^{l, \text{ext}}_{ij\kappa} = e^{2} \sum_n \int_{k} [d{\bf k}] \: j^{l}_{i,n}\frac{\tau^{2}}
{\hbar^{2}}\frac{\partial^{2} f^{(0)}_n}{\partial k_{j}\partial{k}_{\kappa}}.
\end{equation}
In our case the indices $j$ and $k$ are $x$. So dropping the indices $j, k$ and
assuming the relaxation time $\tau $ is independent of energy, 
we can rewrite this equation as,
\begin{equation*}
    \Gamma^{l, \text{ext}}_{i,\pm} = \frac{e^{2}\tau^2}{(2\pi\hbar)^2}  
    \int_{k} d^2k \: \langle \pm|\hat{j}^{l}_{i}|\pm\rangle
\: \frac{\partial^{2} f^{(0)}}{\partial k_{x}^2}.
\end{equation*}

The derivative of the the Fermi-Dirac distribution at very low temperature is can be expressed as
\begin{equation}
    \frac{\partial f^{(0)}}{\partial \varepsilon^{\pm}} \simeq -\delta(\varepsilon^{\pm}
- \varepsilon_F) = -\frac{\delta(k
- k_F^{\pm})}{\left|\frac{\partial \varepsilon^{\pm}}{\partial k_F^{\pm}}\right|}. 
\end{equation}
Putting this in the above integral we get,
\begin{equation}
    \Gamma^{l, \text{ext}}_{i,\pm} = \frac{e^{2}\tau^2}{(2\pi\hbar)^2}  \int_{k} d^2k \: \frac{\p}{\p k_x}(\langle \pm|\hat{j}^{l}_{i}|\pm\rangle)\: \frac{\p \varepsilon^{\pm}}{\p k_{x}}\frac{\delta(k
- k_F^{\pm})}{\left|\frac{\partial \varepsilon^{\pm}}{\partial k_F^{\pm}}\right|},
\end{equation}
where integration by parts is used to shift one of the $k_x$ derivatives on to the current density. The integral is converted to polar coordinates, where the $k$-integral is killed by the delta function and the $\phi$-integral is straight-forward. 

\bibliography{Manuscript}

\end{document}